\documentclass[10pt]{article}
\usepackage{amsmath,amssymb,bm,cite,mathrsfs}
\usepackage{graphicx}

\begin{document}
\title{Polymer transport in random flow}
\author{A. Celani\footnote{CNRS, INLN, 1361 Route des Lucioles, 06560
Valbonne, France}
\footnote{CNRS UMR 6202, Observatoire de la C\^ote d'Azur,
B. P. 4229, 06304 Nice Cedex 4, France}\hspace{0.15cm}, 
S. Musacchio\footnote{INFM, Dipartimento di Fisica, Universit\`a di Roma ``La Sapienza'',
P.le A. Moro 2, 00185 Roma, Italy
}\hspace{0.15cm}, 
and
D. Vincenzi\footnotemark[2]
\footnote{D\'epartement de Math\'ematiques, 
Universit\'e de Nice-Sophia Antipolis, Parc Valrose, 06108 Nice Cedex 2, 
France}
}
\date{\today}

\begin{titlepage}
\maketitle
\begin{list}{}{\labelsep=0.7cm\labelwidth=5cm\leftmargin=5cm}
\item[Suggested running head:] Polymer transport in random flow 
\item[Mailing address:] Dario Vincenzi\\
CNRS, Observatoire de la C\^ote d'Azur\\
Bd. de l'Observatoire B. P. 4229\\
06304 Nice Cedex 4, France
\item[Telephone number:] +33-4-92003172
\item[Fax number:] +33-4-92003121 
\item[E-mail address:] Dario.Vincenzi@obs-nice.fr
\end{list}
\vspace*{1cm}
To be published in the {\it Journal of Statistical Physics}
\thispagestyle{empty}
\end{titlepage}

\begin{abstract}
The dynamics of polymers in a random smooth flow
is investigated in the framework of the Hookean dumbbell model.
The analytical expression
of the time-dependent probability density function of polymer elongation
is derived explicitly for a Gaussian, rapidly changing flow.
When polymers are in the coiled state the pdf reaches a stationary state 
characterized by power-law tails both for small and large arguments
compared to the equilibrium length. The characteristic relaxation time is
computed as a function of the Weissenberg number.
In the stretched state the pdf 
is unstationary and exhibits multiscaling. 
Numerical simulations for the two-dimensional Navier-Stokes flow 
confirm the relevance of theoretical results 
obtained for the $\delta$-correlated model.
 \end{abstract}


\section{Introduction}
A polymer is a long, repeating chain formed through the linkage of many
identical smaller molecules called monomers.
Understanding the behavior of a single polymeric chain passively
advected by a 
turbulent flow is the first step in the study of polymer transport.
At equilibrium a polymer molecule coils itself up into
a ball of radius $R_0$.
When placed in a non-homogeneous flow, the polymer deforms and stretches;
if the number of monomers composing the molecule is large,
the end-to-end distance of the polymer, $R$,
can become much greater than the equilibrium size $R_0$.
A sharp transition from the coiled state to the strongly stretched one
is observed when the strain rate exceeds a critical value~\cite{DeGennes,
Perkins}.

Two competing effects determine the
deformation of a polymer: the stretching by the velocity
gradients and the elastic relaxation towards the equilibrium shape.
Theoretical and numerical results suggest that
a wide range of scales exists 
where the elasticity of the polymer can be described by Hooke's 
law~\cite{Hatfield}. 
In the absence of flow the configuration of the polymer
converges exponentially to the equilibrium shape: $R\approx R_0\, e^{-t/\tau}$,
where $\tau$ is the characteristic time of the slowest oscillation mode.
In contrast, a non-homogeneous velocity field can stretch the polymer
and elongate it.
The characteristic value of velocity
gradients is determined by the maximum Lyapunov exponent $\lambda$, which is
the average logarithmic growth rate of fluid particle separations
(see, e.g., ref.~\cite{Falkovich}).

In random flows the ratio of the relaxation time to
the characteristic stretching one is determined by 
the Weissenberg number $\text{Wi}\equiv\lambda\tau$.
When $\text{Wi}<1$ 
contraction is faster and the extension of the molecule converges definitively
to the equilibrium size; the polymer, therefore, is in the coiled state.
When $\text{Wi}>1$
the elastic force is weak and the velocity gradients can deform 
considerably the molecule. In this case the dynamics of the polymer
depends strongly 
on the properties of the advecting flow. Balkovsky {\it et al.} have shown
that in random flows the threshold $\text{Wi}=1$ marks
the coil-stretch transition and
polymers are highly stretched when Wi exceeds 
one~\cite{Balkovsky00,Balkovsky01}. 
Only very recently has the
coil-stretch transition in random flows been observed experimentally 
by Gerashchenko {\it et al.}~\cite{Gerashchenko}.
It should be however noted that
there can be particular flows where the coil-stretch transition never occurs.
This generally happens when a strong rotation prevents the molecules from 
getting aligned with the stretching direction~\cite{Lumley72}.

When polymers are highly stretched the coupling between 
elastic and kinetic degrees of freedom should be taken into account.
The feedback on the advecting flow is then modeled by an additional term in 
the stress tensor appearing in the Navier-Stokes equations. The most 
important effect of the back reaction on the velocity field is probably
the drag reduction (see, e.g., ref.~\cite{Gyr}).
The passive approach, however, is justified when the stress due to polymers is
much smaller than the viscous stress, which can be estimated as
$\nu_v\lambda$, $\nu_v$ being
the kinematic viscosity of the fluid~\cite{Balkovsky00,Balkovsky01}.

We focus on passive polymer transport: the velocity field is given 
and reaction effects on the advecting flow are neglected.
We study the coil-stretch transition when the
extension of polymers  
is much smaller than their maximum length $R_{max}$ and the correlation
time of the velocity field is negligible. 
The polymer is modeled by a Hookean elastic dumbbell and the
turbulent advecting flow belongs to the
Batchelor-Kraichnan statistical ensemble~\cite{Batchelor,Kraichnan68}. 
The same model was previously investigated by Chertkov~\cite{Chertkov_00} and 
Thiffeault~\cite{Thiffeault_03}. 
Chertkov derived an approximate
equation for the right tail of the 
probability density function (pdf) of the end-to-end 
separation of the dumbbell~\cite{Chertkov_00}.
Thiffeault obtained asymptotic results for the stationary solution of this equation in three 
dimensions~\cite{Thiffeault_03}. We derive 
the Fokker-Planck equation for the complete pdf of the end-to-end 
distance of the polymer in general dimension $d$. 
We thus obtain the exact stationary form and the 
finite-time behavior of the pdf of the elongation. Finally, we compare 
theoretical results 
with numerical simulations for the two-dimensional
Navier-Stokes flow.

The rest of this paper is organized as follows. Section 2 is devoted 
to the description of the model. In Section 3 we derive the 
Fokker-Planck equation for the probability density function 
of the extension $R$. The stationary solution and the finite time
pdf are discussed in Section 4 and in Section 5 respectively.
In Section 6 we discuss the results of numerical simulations
of polymer dynamics in the two-dimensional Navier-Stokes flow.
Section 7 is reserved to conclusions.

\section{Elastic dumbbell model}

A simple description of the complex structure of polymers is provided by
Rouse's model, which represents a polymer as a chain of $N$ beads connected 
by Hookean springs~\cite{Rouse}. 
The dynamics can be solved by decomposing the motion of the molecule
into a set of linear normal modes. A typical relaxation
time is associated with the amplitude of each mode.
In many physical applications the dynamics of the polymer is dominated by 
the fundamental mode, whose relaxation time is of the order of $\zeta/k$,
where $\zeta$ is the friction coefficient of the polymer and $k$ is the 
spring constant.

\begin{figure}[h]
\centering
\includegraphics[height=5cm]{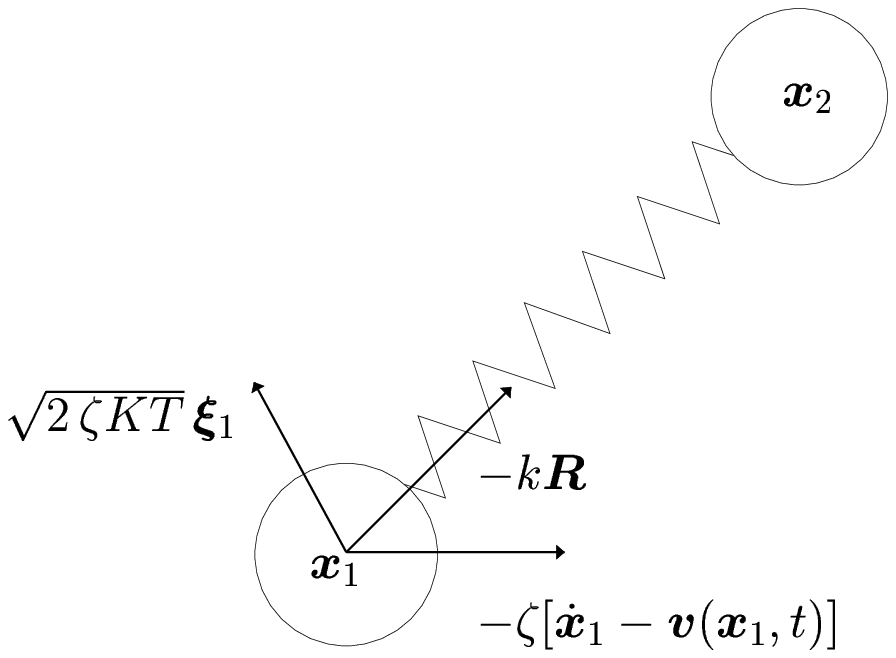}
\caption{}
\label{fig:1}
\end{figure}

The elastic dumbbell model is a simplification of Rouse's model.
Only the fundamental oscillation mode is taken into account and the 
elasticity of the polymer is modeled by a single spring connecting the ends
of the molecule. 
Two beads of radius $l$ are joined at their centers
by a linear zero-length spring of stiffness $k$.
For a chain of $N$ identical monomers of length $L$ the spring constant is
$k=3KT/(N-1)L^2$, where $K$ is the Boltzmann constant and $T$ the temperature%
~\cite{Bird_I}. The positions of the beads are defined by the vectors
$\bm x_1$ and $\bm x_2$; the two beads represent the ends of the polymer and
their separation $\bm R=\bm x_2-\bm x_1$ determines the elongation of the 
molecule (fig.~\ref{fig:1}). The idea of using a dumbbell to simulate a 
polymer dates back to Kuhn~\cite{Kuhn}. For the sake of
completeness, we review the 
derivation of the governing equation for the bead connector 
(see, e.g., ref.~\cite{Bird_II}).

The hydrodynamic drag force that one bead experiences due to its relative
motion to the fluid is given by Stokes' law: 
$-\zeta[\dot{\bm x}_i-\bm v(\bm x_i,t)]$, where $\dot{\bm x}_i$ is the
velocity of the bead $i$ and $\bm v(\bm x_i,t)$ the velocity of the fluid 
at the position of the bead\footnote{
In general we ought to take into account the perturbation of the velocity 
field due to the presence of the other bead and Stokes' law should be
modified as follows:
$-\zeta[\dot{\bm x}_i-\bm v(\bm x_i,t)-\bm v'(\bm x_i,t)]$, 
$\bm v'$ being the perturbation. To a first approximation we neglect this
correction (referred to as ``hydrodynamic interaction'') which would
cause the largest relaxation time to decrease as a result of 
the cooperative motion among the beads~\cite{Bird_II}.}. 
The friction coefficient is $\zeta=
6\pi l\eta_v$, where $\eta_v$ denotes the dynamical viscosity of the solvent.

The size of the beads is assumed to be small enough for their dynamics to be
influenced by thermal noise.
This is modeled by a Gaussian process 
$\sqrt{2\zeta KT}\,\bm\xi_i$ such that~\cite{Risken_89} 
\begin{equation}
\label{eq:xi}
\langle\bm\xi_i(t)\rangle=0
\qquad \langle\xi^k_i(t)\,\xi_j^\ell(t')\rangle=
\delta_{ij}\delta^{k\ell}\delta(t-t')\qquad (i,j=1,2)
\end{equation}
Owing to molecular noise
the equilibrium size of the dumbbell is not zero, but it can be estimated
as $R_0=\sqrt{KT/k}$ by considering 
the elastic energy of the polymer,
$E=kR^2/2$, and its equilibrium value at temperature $T$, $E=KT/2$.

Taking into account the elastic force, the hydrodynamic drag, and
the thermal noise, we obtain the dynamical equations for the beads:
\begin{equation}
\label{eq:beads}
\begin{array}{c}
m\ddot{\bm x}_1=
-k\left(\bm x_1-{\bm x}_2\right)-
\zeta\,\left[\dot{\bm x}_1-\bm v(\bm x_1,t)\right]
+\sqrt{2\,\zeta KT}\,\bm\xi_1\\[0.3cm]
m\ddot{\bm x}_2=
-k\left(\bm x_2-{\bm x}_1\right)-
\zeta\,\left[\dot{\bm x}_2-\bm v(\bm x_2,t)\right]
+\sqrt{2\,\zeta KT}\,\bm\xi_2
\end{array}
\end{equation}
In physical applications the elongation of the polymer, however large it is,
never reaches the viscous scale
of the turbulent flow, where dissipation and advection balance~\cite{%
Lumley}. Below this scale velocity fluctuations are attenuated 
by the viscosity
and the flow is smooth. Thus, during its evolution
the polymer always moves in a regular velocity field, which at the scale
$R$ 
can be approximated by a uniform gradient flow~\cite{Batchelor}:
\begin{displaymath}
\bm v(\bm x_2,t)=\bm v(\bm x_1,t)+(\bm x_2 -\bm x_1)\cdot\nabla\bm v(t)
\end{displaymath}
Neglecting inertial effects\footnote{
The masses of the beads can be eliminated in the limit where
the time scale of Brownian
fluctuations of the velocity, $m/\zeta$, is much smaller than the relaxation
time of the dumbbell, $\zeta/2k$. This corresponds to
considering over-damped oscillations~\cite{Schieber}.}, we finally 
derive from~\eqref{eq:beads} a stochastic ordinary
differential equation for the end-to-end separation of the polymer~\cite{Bird_II}:
\begin{equation}
\label{eq:dumbbell}             
\dot{\bm R}=(\bm R\cdot\nabla) \bm v-\frac{\bm R}{\tau}+
\sqrt{\frac{2R_0^2}{\tau}}\;\bm \xi
\end{equation}
where $\tau=\zeta/2k$ is the relaxation time and $\bm\xi\equiv
(\bm\xi_2-\bm\xi_1)/\sqrt{2}$ has the same properties as $\bm\xi_1$ and
$\bm\xi_2$ [see eq.~\eqref{eq:xi}]. Equation~\eqref{eq:dumbbell} governs the
dynamics of the elongation of a polymer transported by a non-homogeneous flow:
the velocity gradients stretch the 
molecule, which reacts elastically and attempts to recover the equilibrium
spherical shape. Thermal fluctuations of the solvent are modeled by
the white noise $\sqrt{2R_0^2/\tau}\,\bm\xi$.

It should be noted that the infinitely extensible  dumbbell is a 
good approximation of a polymer
only when the extension of the molecule is much smaller
than its maximum length $R_{max}$. 
For very large elongations ($R\lesssim R_{max}$)
the relaxation time becomes
a function of $R$ and the resulting elastic force is nonlinear~\cite{Hatfield}.
In more realistic dumbbell models, 
Hooke's law is replaced by a force diverging as the
size of the polymer approaches  $R_{max}$
(for the description of the dynamics of a
FENE dumbbell transported by the Batchelor-Kraichnan flow
see refs.~\cite{Thiffeault_03,Vincenzi}). 
Our results thus apply to  the range $R\ll R_{max}$.

\section{Fokker-Planck equation}

The probability density function of the elongation,
$P(\bm R,t)$, obeys the Fokker-Planck equation associated with
eq.~\eqref{eq:dumbbell} (see, e.g., refs.~\cite{Risken_89,Gardiner_85})
\begin{equation}
\label{eq:FP rumore}
\frac{\partial P}{\partial t}+
\frac{\partial }{\partial R^i}
\left(R^j \nabla^j v^i - \frac{R^i}{\tau}\right)P=
\frac{R_0^2}{\tau}\frac{\partial^2 P}{\partial R^i \partial R^i}
\end{equation}
where summation over repeated indices is understood.
In the passive regime the turbulent advecting flow has to
be specified. Within the context of Kraichnan's model, $\bm v$
is a Gaussian random field with zero mean and correlation function~\cite{%
Kraichnan68}
$$
\langle v^i(\bm x,t)\, v^j(\bm x +\bm r,t')\rangle=
{\mathscr D}^{ij}(\bm r)\,\delta(t-t')
$$
The velocity field is by definition
statistically homogeneous in space and stationary in time. The non-realistic
property of the Kraichnan ensemble is the $\delta$-correlation in time,
which defines a stochastic process with no memory. Nevertheless, 
the absence of time correlation in the advecting flow
yields closed equations for the correlation functions of the transported field.
The study of this model
brought important results in turbulent transport theory; these can be
interpreted as the limit of real behaviors 
as the velocity correlation time 
tends to zero (see ref.~\cite{Falkovich} for a review).
A complete definition of Kraichnan's model requires writing 
the covariance tensor ${\mathscr D}^{ij}(\bm r)$ explicitly. If we impose 
incompressibility, smoothness, and
statistical invariance under parity and rotations,
${\mathscr D}^{ij}(\bm r)$ takes the form~\cite{Yaglom}
$$
{\mathscr D}^{ij}(\bm r)=D_0\delta^{ij}-D_1[(d+1)\delta^{ij}r^2-2r^ir^j]
$$
where the constant $D_0$ is the eddy diffusivity of the flow, 
$D_1$ determines the intensity of the fluctuations of the
velocity, and $d$ is the physical dimension of the flow ($d=2,3$ in practical
applications). Consequently, the two-time 
correlation of the velocity gradients takes the form
\begin{equation}
\label{eq:gradienti}
\langle \nabla^i v^j (t)\, \nabla^k v^\ell (t')\rangle=
2D_1[(d+1)\delta^{ik}\delta^{j\ell}-\delta^{ij}\delta^{k\ell}-\delta^{i\ell}\delta^{jk}]
\,\delta(t-t')
\end{equation}
It is worth noticing that when $\bm v$ is a Gaussian white noise 
the advection term in
eq.~\eqref{eq:dumbbell} defines a multiplicative stochastic process
like the ones considered in refs.~\cite{Schenzle_PhysLett,Schenzle_PhysRev,
Graham_80,Graham_82}.

By averaging eq.~\eqref{eq:FP rumore}
over realizations of the velocity field, we obtain the 
evolution equation for the average probability density function
$\widehat{P}(\bm R,t)\equiv\langle P(\bm R,t)\rangle_v$:
\begin{multline}
\label{eq:hat}
\frac{\partial \widehat{P}}{\partial t}
-D_1\left[(d+1)\frac{\partial}{\partial R^i}R^j\frac{\partial}{\partial R^i}R^j
-\frac{\partial}{\partial R^i}R^j\frac{\partial}{\partial R^j}R^i
-\frac{\partial}{\partial R^i}R^i\frac{\partial}{\partial R^j}R^j\right]\widehat{P}
-\\
\frac{1}{\tau}\frac{\partial}{\partial R^i}R^i \widehat{P}
=\frac{R_0^2}{\tau}\frac{\partial^2\widehat{P}}{\partial R^i\partial R^i}
\end{multline}
The term 
$\langle \nabla^jv^i \widehat{P}\rangle_v$ can be computed by Gaussian 
integration by parts~\cite{Furutsu,Novikov,Frisch}
and by using eq.~\eqref{eq:gradienti}.

To analyze the coil-stretch transition, we can restrict our study 
to the pdf of the norm of $\bm R$, $p(R,t)\equiv
\int\widehat{P}(\bm R,t)R^{d-1}d\Omega$, where $d\Omega$ denotes integration
over angular variables. All the differential operators appearing in 
eq.~\eqref{eq:hat} are scalar; therefore it is easily shown that $p(R,t)$ 
obeys the one-dimensional Fokker-Planck equation
\begin{equation}
\label{eq:FP}
\frac{\partial p}{\partial t}=-\frac{\partial}{\partial R}
\left[{\cal K}_1(R)p\right]
+\frac{\partial^2}{\partial R^2}\left[{\cal K}_2(R)p\right]
\end{equation}
where ${\cal K}_1$ and ${\cal K}_2$ are
the drift and diffusion coefficients respectively:
\begin{gather}
\label{eq:coefficienti}
{\cal K}_1(R)=a(1-q)R+\dfrac{b(d-1)}{R}\qquad\qquad
{\cal K}_2(R)=aR^2+b\\[0.5cm]
\label{eq:esponente}
a=\dfrac{\Delta}{2}  \qquad \qquad b=\dfrac{R_0^2}{\tau} \qquad\qquad
q=\dfrac{2}{\Delta}\left(\dfrac{1}{\tau}-\lambda\right)
\end{gather}
The drift and diffusion coefficients
are time-independent due to the stationarity of the advecting velocity field.
The parameter $\lambda=D_1 d(d-1)$ coincides with the maximum Lyapunov
exponent of the 
Batchelor-Kraichnan model, while $\Delta=2\lambda/d$ 
is the variance
of the asymptotic distribution of Lyapunov exponents and is related
to the correlation of velocity gradients~\cite{Kraichnan,LeJan,LeJan_84,Son}
(see also ref.~\cite{Falkovich}). 
The constant $q$ depends on the Weissenberg number $\text{Wi}=\lambda\tau$:  
$q=d(1-\text{Wi})/\text{Wi}$.
While the other parameters $a$, $b$, $d$ are always positive,
the sign of $q$ is positive for $\text{Wi}<1$ and
negative in the opposite case.  

Equation~\eqref{eq:FP} should 
 be supplemented by an initial condition $p(R,0)$
and by suitable boundary conditions on the interval $(0,\infty)$.
To this aim the Fokker-Planck equation may be rewritten in the form
of a probability conservation equation 
\begin{equation}
\label{eq:corrente}
\frac{\partial p}{\partial t}+\frac{\partial J}{\partial R}=0
\end{equation}
where the functional
\begin{equation*} 
J[p(R,t)]\equiv{\cal K}_1(R)\,p(R,t)-\frac{\partial }{\partial R}
\left[{\cal K}_2(R)\,p(R,t)\right]
\end{equation*}
is the probability current~\cite{Risken_89,Gardiner_85,Stratonovich_63}.
The boundary conditions may then be expressed in terms of $J(R,t)$.
The normalization of the pdf for all times~$t$ implies necessarily:
$J(0,t)=\lim_{R\to\infty}J(R,t)$ $\forall \,t>0$.
Here we consider a little stronger 
conditions, usually imposed to the solution of the Fokker-Planck equation
on an infinite domain: we search for a solution such that the 
associated probability current vanishes both in zero and at infinity:
\begin{equation}
\label{eq:contorno}
J(0,t)=
\lim_{R\to\infty}J(R,t)=0
\qquad \quad\forall\, t>0
\end{equation}
These conditions (called reflecting boundary conditions)
correspond to requiring that there is no flow of probability
through the boundaries of the domain of 
definition~\cite{Risken_89,Gardiner_85,Stratonovich_63}.

When $R_0$ is set to zero eq.~\eqref{eq:FP} reduces to the equation 
obtained by Chertkov by replacing the norm of the end-to-end vector 
with its maximum component~\cite{Chertkov_00}. 
This approximate equation gives the
right tail of the pdf of the elongation. Starting from the equation
derived by Chertkov,
Thiffeault studied the large-$R$ tail of the stationary pdf in three
dimensions~\cite{Thiffeault_03}.
Our equation~\eqref{eq:FP} governs the time evolution of the complete pdf 
of the elongation in general dimension $d$.
In the 
following sections we shall  compute 
the exact stationary form and the finite-time 
behavior of the pdf of the elongation for a Hookean dumbbell  
transported by the Batchelor-Kraichnan flow.

\section{Stationary solution}

Since the drift and diffusion coefficients do not depend on time, 
$p(R,t)$ might relax to a stationary 
distribution $p_0(R)$ independent of time and of the initial condition.
Under reflecting boundary conditions~\eqref{eq:contorno} the probability
current associated with $p_0(R)$ should vanish identically.
The stationary pdf may then be sought in the form~\cite{Risken_89,Gardiner_85,Stratonovich_63}
\begin{equation}
\label{stazionaria}
p_0(R)=\frac{C}{{\cal K}_2(R)}\,
\exp\left[\int_{R_1}^R\frac{{\cal K}_1(y)}{{\cal K}_2(y)}\, dy\right]
\end{equation}
where the constant $C$ and the lower integration limit $R_1$ are fixed 
by the normalization condition.

\begin{figure}[h]
\includegraphics[height=4.4cm]{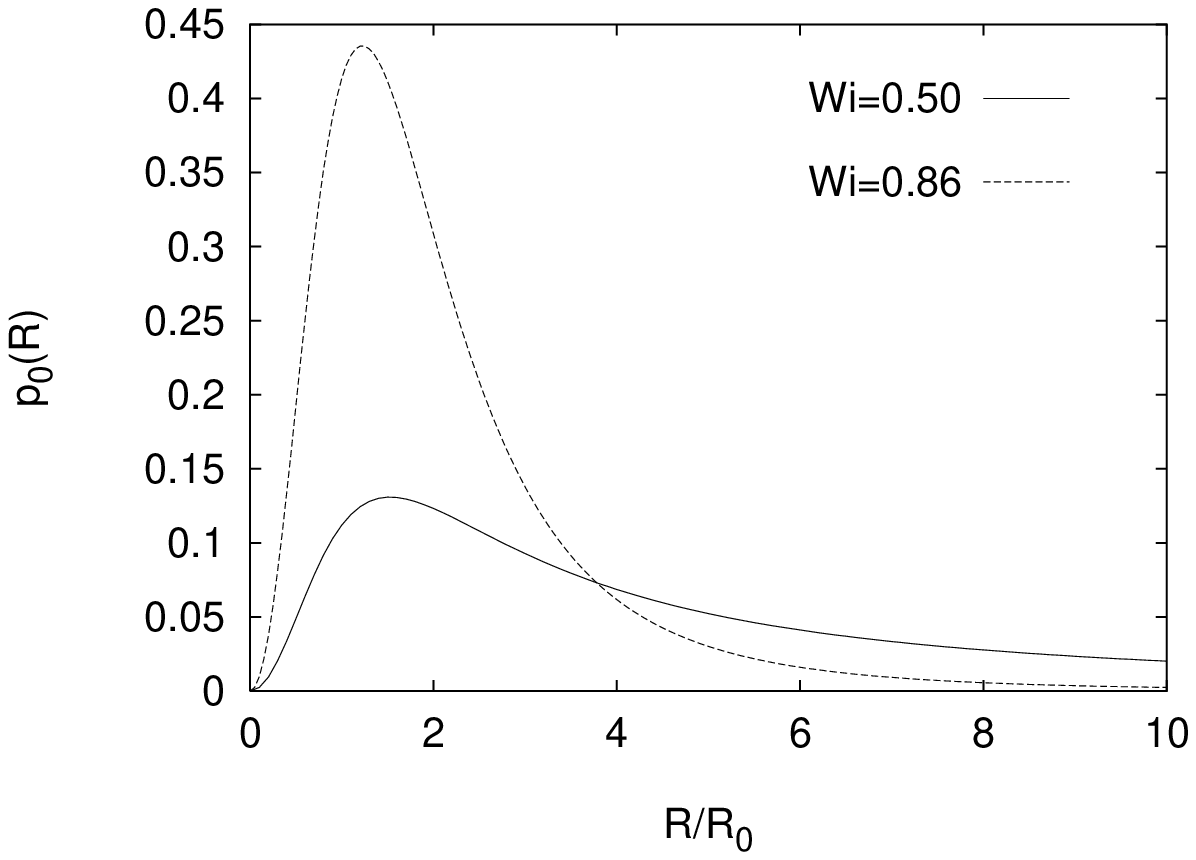}\hfill  
\includegraphics[height=4.4cm]{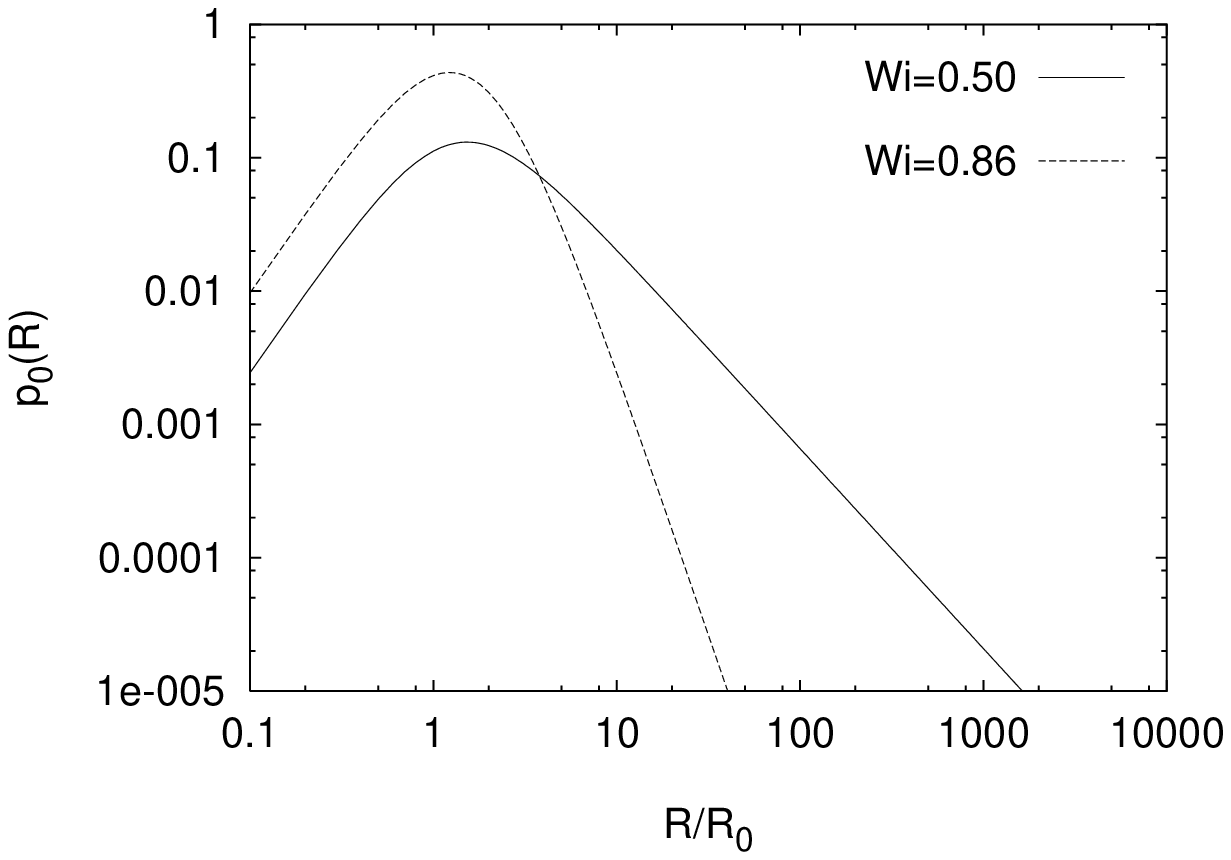}
\caption{}
\label{fig:2}
\end{figure}

Equation~\eqref{eq:FP} has a stationary solution only when $q$ is positive
[$q$ is defined by eq.~\eqref{eq:esponente}].
After inserting definitions~\eqref{eq:coefficienti} and~\eqref{eq:esponente}  
in eq.~\eqref{stazionaria}, we obtain:
\begin{equation*}
p_0(R)=N_0\, R^{d-1}\left(1+\frac{\text{Wi}}{d}
\frac{R^2}{R_0^2}\right)^{-(q+d)/2}\qquad (q>0)
\end{equation*}
$N_0$ is the normalization constant (see ref.~\cite{Gradshteyn} 
formula 3.252.2): 
$$
N_0=2\left(\frac{\text{Wi}}{d}\right)^{d/2}
\frac{\Gamma\left(\frac{q+d}{2}\right)}%
{\Gamma\left(\frac{q}{2}\right)\Gamma\left(\frac{d}{2}\right)}
$$
and $\Gamma$ denotes the Eulerian integral of the second kind.
As the extension of the polymer approaches zero, the stationary pdf vanishes as
\begin{equation}
p_0(R)=O(R^{d-1}) \qquad \text{as}\ R\to 0 
\label{eq:codasx}
\end{equation}
Small values of $R$ are indeed expected when the stretching term is negligible 
and eq.~\eqref{eq:dumbbell} reduces to an Ornstein-Uhlenbeck 
process~\cite{Risken_89}.
The stationary pdf has a maximum at $R\approx R_0$ and displays an algebraic
tail for large $R$:
\begin{equation}
p_0(R)=O(R^{-1-q}) \qquad \text{as}\  R\to \infty
\label{eq:codadx}
\end{equation}
This property has been predicted by Balkovsky {\it et al.} for a
generic random velocity field~\cite{Balkovsky00} and has been confirmed 
by numerical simulations for two-dimensional~\cite{Boffetta03} and 
three-dimensional~\cite{Eckhardt} flows, 
and by experimental studies in random flows~\cite{Gerashchenko}.
If $q>0$ (and so $\text{Wi}<1$), the normalization integral 
$\int_0^\infty p_0(R)\,dR$ is dominated by small $R$: most of the
polymers have a linear size of the order of $R_0$ (fig.~\ref{fig:2}). 
On the contrary, for $q<0$ (or $\text{Wi}>1$) 
the normalization integral does not converge. This means 
that a stationary pdf does not exist: 
most of the molecules are highly 
stretched and the passive approach is no longer appropriate for the 
Hookean dumbbell model. Thus, $q=0$ ($\text{Wi}=1$) is the threshold for the
coil-stretch transition~\cite{Balkovsky00,Balkovsky01,Gerashchenko}.

The characteristic exponent $q$ defined in eq.~\eqref{eq:esponente}
coincides with the exponent estimated by Balkovsky {\it et al.}
in the neighborhood of the transition ($\tau\approx\lambda^{-1}$).
Indeed, they considered a quadratic approximation of the
Cram\'er function~\cite{Frisch} 
associated with large deviations of the Lyapunov exponents
of the flow, and
for a velocity field with the Batchelor-Kraichnan statistics, the Cram\'er 
function is exactly quadratic for all $\lambda$~\cite{Kraichnan}.

For a fixed positive $q$ only a finite number of moments of the stationary pdf
converge. This is due to the presence of highly stretched polymers.
The number of diverging moments grows with decreasing $q$ and large 
elongations become more and more probable.
The $m$-th order moment can be computed explicitly as
\begin{equation}
\label{eq:Rm}
\langle R^m\rangle\equiv
\int_0^\infty R^m p_0(R)\, dR=\left(\frac{\text{Wi}}{d}\right)^{-\frac{m}{2}}
\frac{\Gamma\left(\frac{d+m}{2}\right)\Gamma\left(\frac{q-m}{2}\right)}
{\Gamma\left(\frac{d}{2}\right)\Gamma\left(\frac{q}{2}\right)}\qquad (m<q)
\end{equation}
This latter expression is in accordance with the approximation deduced
by Thiffeault for $d=3$~\cite{Thiffeault_03}. 

\section{Finite-time evolution}

The time-dependent solution of the Fokker-Planck equation
can be obtained by separation of variables~\cite{Risken_89,Gardiner_85,Stratonovich_63}. The solution should then be sought in the form  
$p(R,t)=p_\nu(R)\,e^{-\nu t}$ and the study of eq.~\eqref{eq:FP} reduces
to the eigenvalue problem
\begin{equation}
\label{autovalori}
\mathscr{L}_{\text{\sc fp}}\, p_\nu=-\nu p_\nu
\end{equation}
where $\mathscr{L}_{\text{\sc fp}}$ is the Fokker-Planck differential operator
$$
\mathscr{L}_{\text{\sc fp}} \cdot \equiv -\frac{\partial}{\partial R}\left({\cal K}_1(R)\cdot\right)
+\frac{\partial^2}{\partial R^2}\left({\cal K}_2(R)\cdot\right)
$$
The general solution of eq.~\eqref{eq:FP} can then be written as an 
expansion in terms of the eigenfunctions $p_\nu$ where the coefficients
are fixed by the initial condition.

The eigenvalue equation associated with the Fokker-Planck 
equation~\eqref{eq:FP} 
can be transformed into a Gauss hypergeometric equation and can therefore
be solved explicitly. The complete solution of the 
eigenvalue problem~\eqref{autovalori} is postponed to appendix~\ref{app:A}. 
Here we describe only the main results.

The form of the spectrum of the operator $\mathscr{L}_{\text{\sc fp}}$
fixes the time behavior of the pdf of $R$.
The spectrum of $\mathscr{L}_{\text{\sc fp}}$ 
in general consists of a discrete as well as a continuous branch:
\begin{itemize}
\item[(a)] for all $q$,
the {\it continuous} branch is the set of all real values $\nu>q^2\Delta/8$;
\item[(b)]
the form of the {\it discrete} branch depends on the value of 
the exponent~$q$:
\begin{itemize}
\item[-]
for a fixed positive $q$, the  discrete branch is bounded from
above by $ q^2\Delta/8$ and consists of a finite number of levels of the form:
$\nu_n=n\Delta(q-2n)$, $n<q/4$.
If $0<q<4$, the only discrete eigenvalue  is $\nu=0$,
which is associated with the 
stationary pdf; with increasing $q$ more levels appear;
\item[-]for negative $q$, 
the spectrum of $\mathscr{L}_{\text{\sc fp}}$ has not a discrete
branch.
\end{itemize}
\end{itemize}
What are the implications for polymer dynamics?

\paragraph{Coiled state}
In the coiled state ($q>0$) the pdf relaxes exponentially 
to a stationary distribution.
For $q$ large enough ($q>4$) the typical relaxation time is the inverse 
of the smallest non-zero eigenvalue of the discrete spectrum, 
{i.e.} $T_1=[\Delta(q-2)]^{-1}$. If $0<q<4$, the convergence to the stationary 
pdf is governed by the continuous branch and the characteristic time is
$T_2=8/(q^2\Delta)$. It is worth noticing  that the relaxation time scale 
depends on the parameter $q$ or equivalently on the 
Weissenberg number. This dependence is linear 
for small Wi and becomes quadratic 
as the system approaches the transition $\text{Wi}=1$ (such behavior is typical
of many nonlinear multiplicative stochastic processes
arising in statistical 
physics~\cite{Schenzle_PhysLett,Schenzle_PhysRev,Graham_80,Graham_82}).

\begin{figure}[h]
\includegraphics[height=4.4cm]{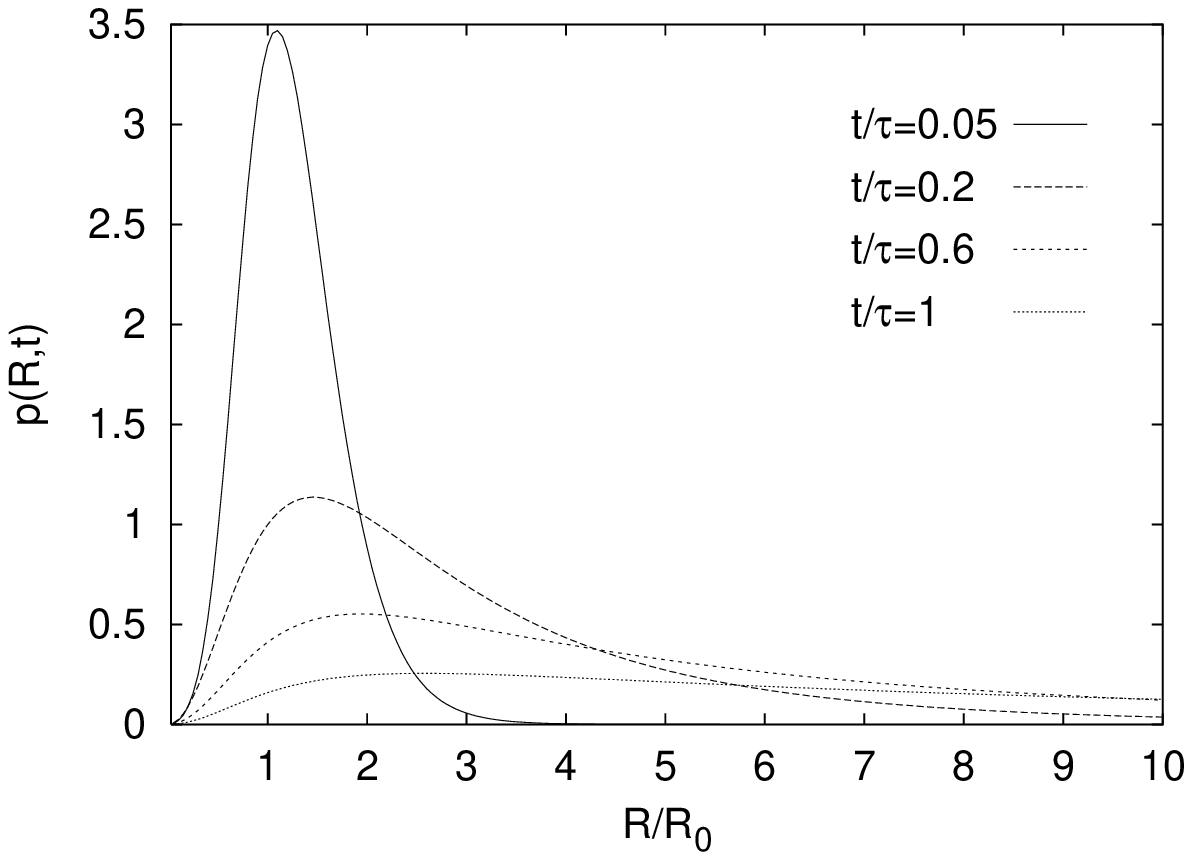}\hfill
\includegraphics[height=4.4cm]{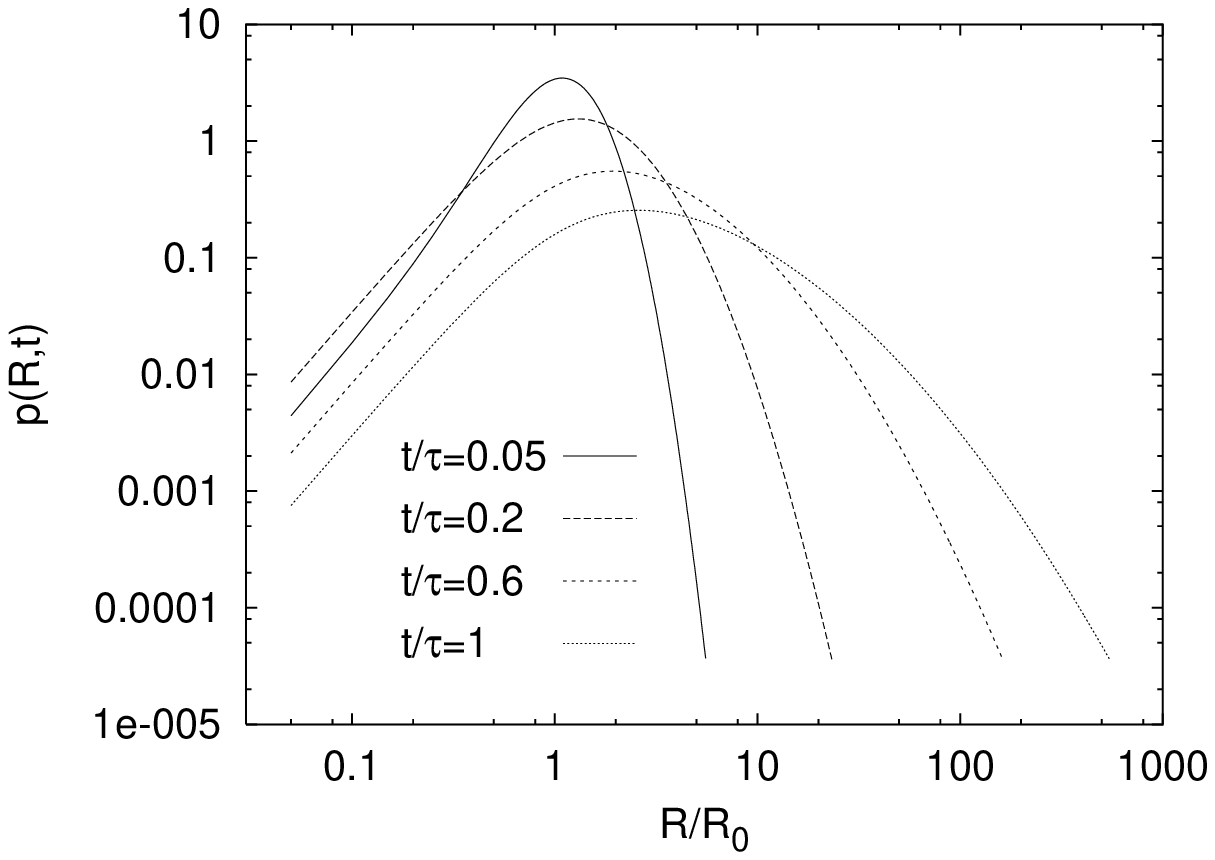}
\caption{}
\label{fig:4}
\end{figure}

\paragraph{Stretched state}
We have already noted that in the stretched state ($q<0$) 
there is no stationary pdf:
as the time increases, the large-$R$ tail of $p(R,t)$ rises,
polymers become more and more elongated and
the passive description is no longer suitable for the Hookean dumbbell model. 
The time evolution of the probability $p(R,t)$ with initial
condition $\delta(R-R_0)$ is shown in fig.~\ref{fig:4}. We have
have computed $p(R,t)$ by means of the expansion in terms of the 
eigenfunctions $p_\nu$ [see eq.~\eqref{eq:expansion} of appendix~\ref{app:A}].

The long time behavior of the pdf of the elongation may also be characterized
by the flatness factors $\langle R^{2m}\rangle/\langle R^2\rangle^m$. 
From eq.~\eqref{eq:FP} the 
moments of the elongation satisfy a hierarchy of ordinary differential 
equations which, for even-order moments, can be easily solved by
Laplace transformation and matrix inversion (see appendix~\ref{app:B}).
The long time behavior of $\langle R^{2m}\rangle$ is exponential with an
$m$-dependent exponent which cannot be deduced from the spectrum of the
Fokker-Planck operator (see, e.g.,  ref.~\cite{Schenzle_PhysRev}):
\begin{equation}
\label{eq:2m}
\langle R^{2m}\rangle=O\big(e^{s_{2m} t/\tau}\big) \qquad (t\to\infty)
\end{equation}
with $s_{2m}=2m [(1+2m/d)\text{Wi}-1]$.
Equation~\eqref{eq:2m} is in accordance with the prediction of 
Chertkov~\cite{Chertkov_00}.  
For a fixed $q$ the large-order flatness factors grow exponentially
in time: the pdf of the elongation is intermittent and deviates
increasingly from a Gaussian distribution.

\section{Polymer dynamics in ``real'' turbulence }

One might wonder whether the conclusions reached for the $\delta$-correlated flow
are of any relevance to real flows. Therefore, we present here results from the 
numerical integration of polymer dynamics, eq.~(\ref{eq:dumbbell}),
with the flow obtained from the two-dimensional Navier-Stokes equations:
\begin{equation}
\partial_t {\bm v} + {\bm v} \cdot \nabla {\bm v} = - \frac{\nabla P}{\rho} 
+ \nu_v \Delta {\bm v} -\alpha {\bm v} + {\bm f}
\label{eq:6.1}
\end{equation}
The velocity field is driven by the large-scale forcing $\bm f$ 
which is modeled by a random zero-mean, statistically 
homogeneous and isotropic, solenoidal vector field.
It counteracts the dissipation due to viscosity $\nu_v$ and friction $\alpha$ 
and allows one to obtain a statistically steady state, 
characterized by a direct enstrophy cascade 
toward small scales \cite{KM80}. 
The ensuing velocity field is smooth and random \cite{BCMV02}.
Numerical integration of eq.~\eqref{eq:6.1} 
has been performed with a standard pseudo-spectral code 
fully dealiased with second-order Runge-Kutta time stepping, 
on a doubly periodic domain of size $L = 2 \pi$. 

\begin{figure}[h]
\centering
\includegraphics[height=5cm]{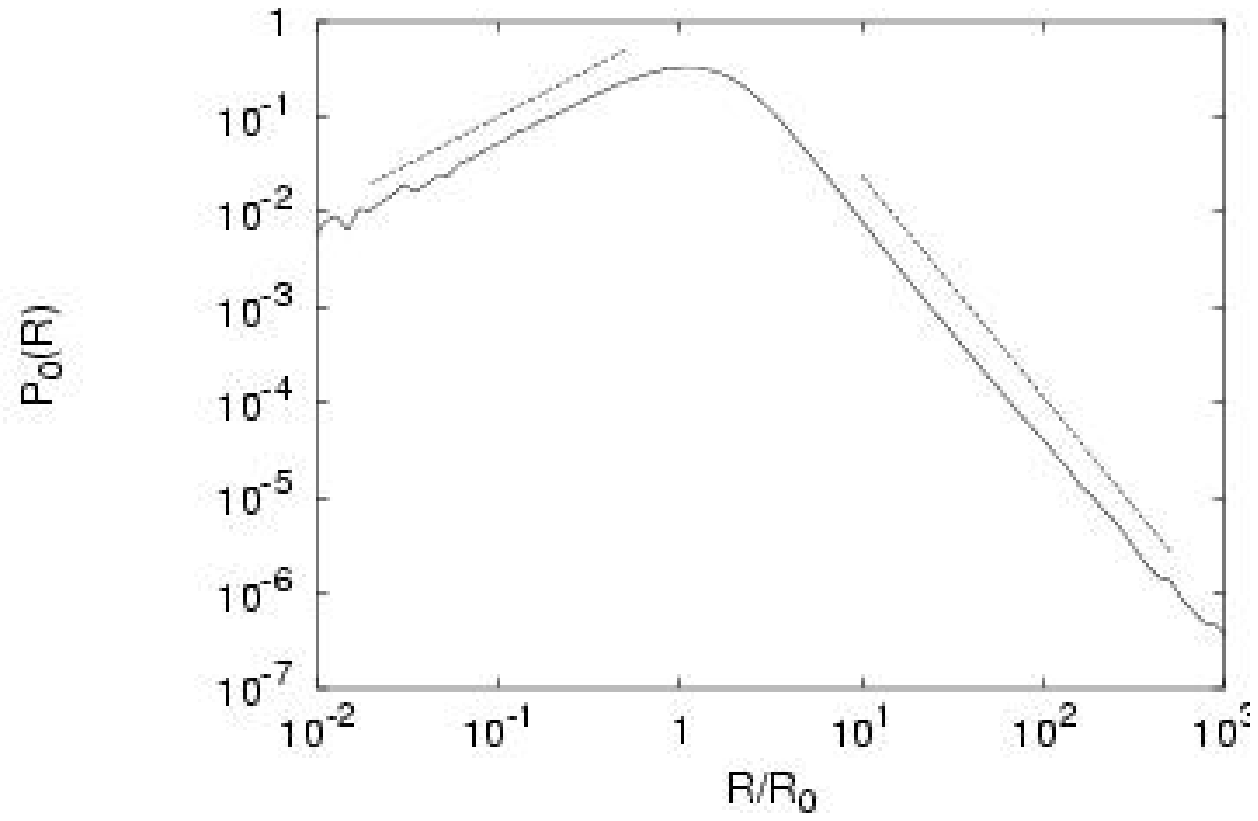}
\caption{}
\label{fig:5}
\end{figure}

\begin{figure}[h]
\includegraphics[height=5cm]{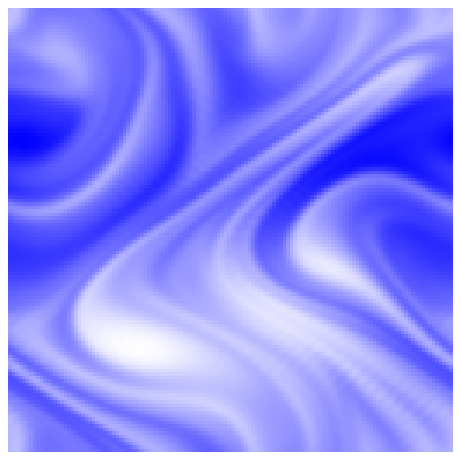}\hfill
\includegraphics[height=5cm,width=5cm]{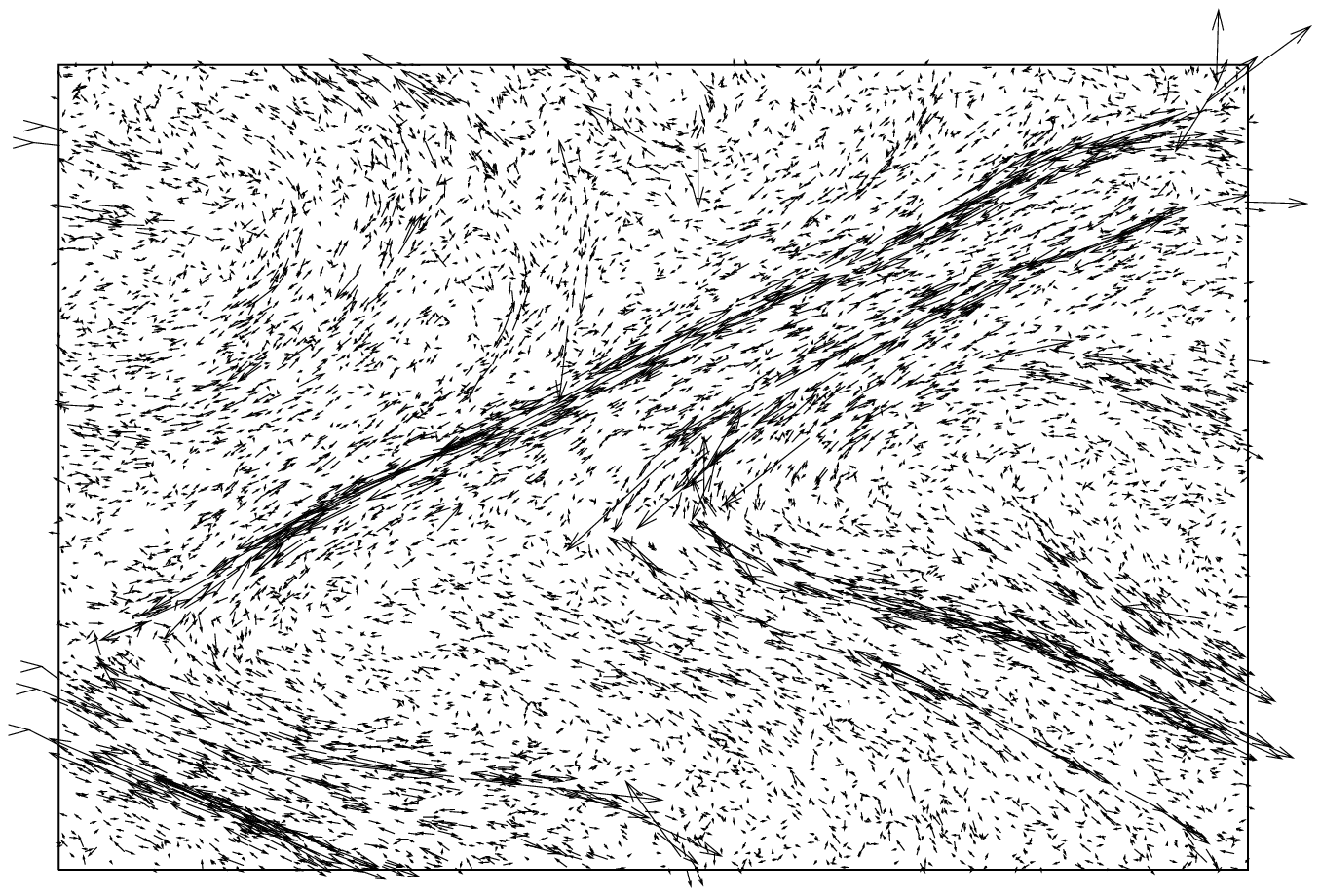}
\caption{}
\label{fig:6}
\end{figure}

\begin{figure}[h]
\centering
\includegraphics[height=5cm]{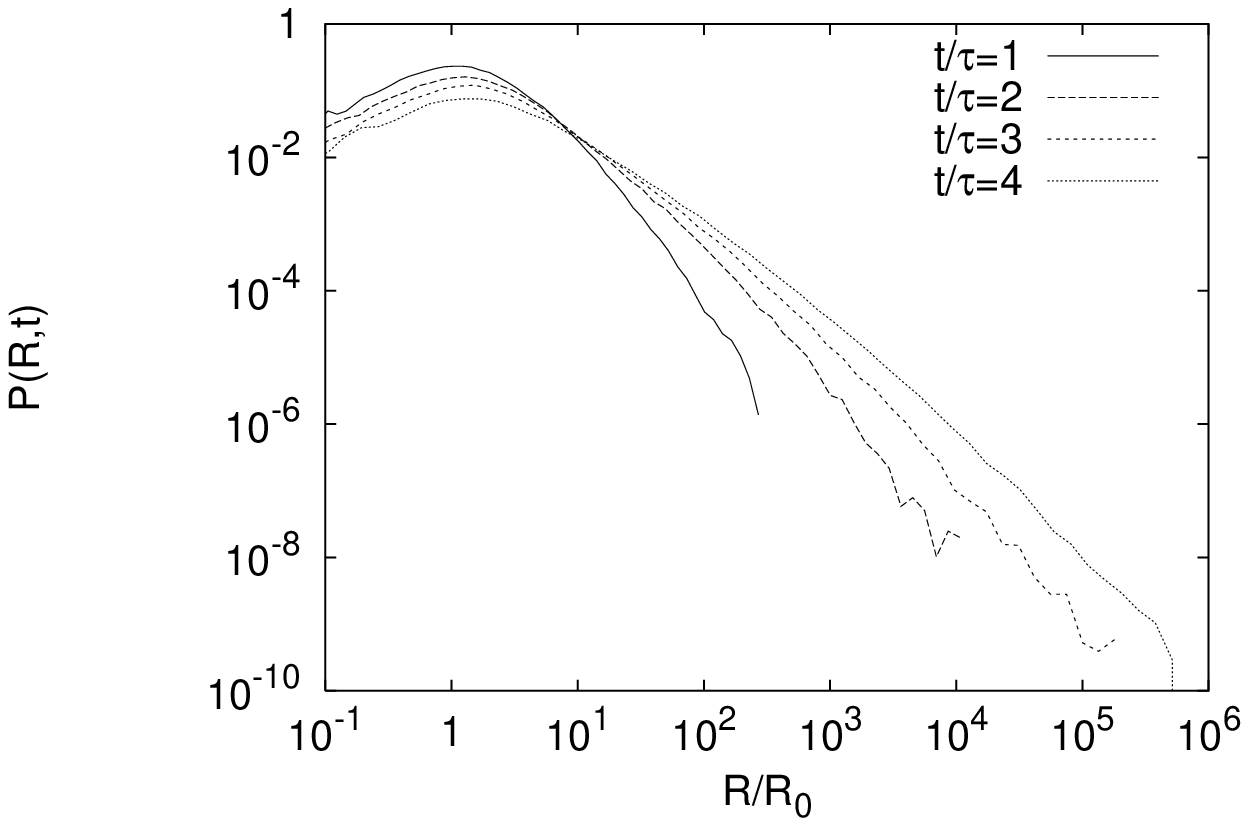}
\caption{}
\label{fig:7}
\end{figure}

In fig.~\ref{fig:5} we show the probability density function for polymer
elongation in the coiled state.
The pdf of $R$ reaches a stationary state characterized by a peak around $R_0$,
and power-law tails both for small and large elongations, in agreement with
the results obtained for the 
Batchelor-Kraichnan case (see eqs.~\eqref{eq:codasx} and
\eqref{eq:codadx}, and fig.~\ref{fig:2}). The inspection of the vector field
of polymer end-to-end separations ${\bm R}$ reveals a strong 
correlation with structures of the velocity field, as displayed in fig.~\ref{fig:6}. 
Above the threshold, the pdfs become unstationary, as shown in fig.~\ref{fig:7},
similarly to what predicted for the short-correlated flow (see fig.~\ref{fig:4}).

Clearly, the basic aspects of polymer stretching by random flows are
well captured by the analysis of the $\delta$-correlated case. 

\section{Conclusions}

We have studied the behavior of polymer molecules 
seeded in a smooth random flow by means of an elastic dumbbell model. 
Asymptotic results for this model were already obtained in 
refs.~\cite{Chertkov_00,Balkovsky00,Balkovsky01,Thiffeault_03}.
The study of 
the $\delta$-correlated flow allows a complete analytical treatment 
and the derivation of the exact pdf of polymer elongation. 
In the coiled state,
 i.e. below the critical Weissenberg number $\text{Wi}=1$, 
the pdf is stationary and characterized
by power laws both for small and for large elongations compared to the
equilibrium length. Above the transition to the stretched state, the pdf
is not stationary anymore and exhibits multiscaling. The same features
are observed  also for polymers in a turbulent two-dimensional Navier-Stokes 
flow: this points to the genericity of the results obtained for the 
Batchelor-Kraichnan flow. 

As a by-product of our analysis, an important and new result is the 
computation of
the typical time of relaxation to the stationary regime in the
coiled state. Two regimes can be identified depending on the value
Weissenberg number:
$$
T/\tau=\begin{cases}
\dfrac{1}{2}\left[1-\left(\dfrac{d+2}{d}\right)\text{Wi}\right]^{-1}
& \text{for $0\le\text{Wi}\le d/(d+4)$ }\\[0.5cm]
\dfrac{4\text{Wi}}{d(1-\text{Wi})^2}
& \text{for $d/(d+4)\le\text{Wi}<1$}.
\end{cases}	
$$
The relaxation time diverges quadratically approaching the critical 
Weissenberg number $\text{Wi}=1$. Therefore, experimental measures near the
coil-stretch transition turn out to be quite delicate because of the
long time needed to reach the stationary regime. For instance, 
a large statistics should be collected to evaluate the scaling exponent $q$
when the Weissenberg number is near the critical value.

The Hookean model is inadequate when polymers become highly elongated.
Our analysis can be extended to nonlinear elastic models
which take into account the finite extensibility of polymers. 
The stationary pdf of the elongation
for a FENE dumbbell in a $\delta$-correlated flow has been 
derived in ref.~\cite{Vincenzi};
the computation of the relaxation time-scale is still an on-going work.
Finally, it would be of great interest to extend the results of the present work
to more realistic models of polymer dynamics,
e.g. the Rouse model \cite{Bird_II}, within the framework of
short-correlated advecting flow.

\appendix

\section{Finite-time pdf}
\label{app:A}
The time-dependent solution of the Fokker-Planck equation
can be obtained by separation of variables through the ansatz:
$p(R,t)=p_\nu(R)\,e^{-\nu t}$
~\cite{Risken_89,Gardiner_85,Stratonovich_63}. 
The study of eq.~\eqref{eq:FP} thus reduces to the eigenvalue problem
\begin{equation}
\label{autovalori_app}
\mathscr{L}_{\text{\sc fp}}\, p_\nu=-\nu p_\nu
\end{equation}
where $\mathscr{L}_{\text{\sc fp}}$ is the Fokker-Planck differential operator
$$
\mathscr{L}_{\text{\sc fp}} \cdot \equiv -\frac{\partial}{\partial R}\left({\cal K}_1(R)\cdot\right)
+\frac{\partial^2}{\partial R^2}\left({\cal K}_2(R)\cdot\right)
$$
Under reflecting boundary conditions $\mathscr{L}_{\text{\sc fp}}$ is symmetric and
negative semi-defined with respect to the scalar product
\begin{equation}
\label{prodotto scalare} 
(f,g)=\int_0^\infty f(R)g(R)[p_0(R)]^{-1}\, dR
\end{equation}
Hence, $\nu$ is real and non-negative. The spectrum of $\mathscr{L}_{\text{\sc fp}}$
in general consists of a 
discrete as well as a continuous branch. If the level 
$\nu=0$ belongs to the spectrum, the Fokker-Planck equation has a 
stationary solution coinciding with the eigenfunction associated with $\nu=0$.
When reflecting boundary conditions are imposed\footnote{%
$J_\nu(0)=\lim_{R\to\infty}J_\nu(R)=0$, $J_\nu$ being the probability current
associated with $p_\nu$.} 
the eigenfunctions $p_\nu$ form an orthogonal set with respect to the scalar 
product~\eqref{prodotto scalare}:
\begin{equation}
\label{eq:ortog}
\int_0^\infty p_{\nu}(R)\,p_{\nu'}(R)[p_0(R)]^{-1} 
\, dR=
\begin{cases}
\delta_{\nu\nu'} & \text{discrete spectrum}\\
\delta(\nu-\nu') & \text{continuous spectrum}
\end{cases}
\end{equation}
Under the further assumption that the set 
$\{p_\nu\}$ is a complete basis,
the time-dependent pdf can be expressed as
\begin{equation}
\label{eq:sviluppo}
p(R,t)=\sum_n A_n \, p_{\nu_n}(R)\, e^{-\nu_n t}+
\int d\nu\, A(\nu) p_\nu(R) e^{-\nu t}
\end{equation}
where the coefficients of the expansion 
are fixed by the initial condition $p(R,0)$:
$$
A_n=\int_0^{\infty}p(R,0)p_{\nu_n}(R)[p_0(R)]^{-1}\, dR
$$
and
$$
A(\nu)=\int_0^{\infty}p(R,0)p_{\nu}(R)[p_0(R)]^{-1}\, dR 
$$
It is worth noticing that the transition probability $p(R,t|\rho,0)$ 
is the particular solution corresponding to the initial condition
$p(R,0|\rho,0)=\delta(R-\rho)$ and can be expanded in the form~\eqref{eq:sviluppo} as
\begin{multline*}
p(R,t|\rho,0)=\\
\sum_n  [p_0(\rho)]^{-1}p_{\nu_n}(\rho) p_{\nu_n}(R)\, e^{-\nu_n t}
+\int d\nu\,  [p_0(\rho)]^{-1}p_{\nu}(\rho)p_\nu(R) e^{-\nu t}
\end{multline*}
The eigenvalue equation associated with the Fokker-Planck 
equation~\eqref{eq:FP} is\footnote{For notational convenience, in this
appendix we use the parameters $a$, $b$,
$d$, $q$ defined in eq.~\eqref{eq:esponente}.}
\begin{multline}
\label{eq:autof}
(aR^2+b)\frac{d^2p_\nu}{dR^2}+
\left[a(q+3)R-\frac{b(d-1)}{R}\right]\frac{dp_\nu}{dR}+\\
\left[a(q+1)+\frac{b(d-1)}{R^2}+\nu\right]p_\nu=0
\end{multline}
By the change of variable $z=-\epsilon R^2$, $\epsilon=a/b$, this latter equation 
can be transformed into a Gauss hypergeometric equation for the new
function $w_\nu(z)=z^{(1-d)/2}p_\nu(z)$. Therefore, 
the solutions of eq.~\eqref{eq:autof} 
fulfilling reflecting boundary conditions take 
the form\footnote{%
The second solution of the Gauss equation scales as 
$R$ in a neighborhood of the origin
for odd $d$ and has a logarithmic singularity at $R=0$ for even $d$~\cite{Goursat}.
Therefore, it does not match reflecting boundary conditions for any $d$.}
\begin{equation}
\label{eq:autofunzioni}
p_\nu(R)=N_\nu\, R^{d-1}F\left(\alpha_\nu,\beta_\nu;\gamma;-\epsilon R^2\right)
\end{equation}
where $N_\nu$ is the normalization constant and 
$F\left(\alpha_\nu,\beta_\nu;\gamma;-\epsilon R^2\right)$
denotes the Gauss hypergeometric function with parameters
\begin{equation}
\label{eq:parametri}
\alpha_\nu=\frac{d}{2}+\frac{q}{4} -\frac{1}{4}\sqrt{q^2-\frac{4\nu}{a}}
\qquad
\beta_\nu=\frac{d}{2}+\frac{q}{4} +\frac{1}{4}\sqrt{q^2-\frac{4\nu}{a}}
\qquad
\gamma=\frac{d}{2}
\end{equation}
From eqs.~\eqref{eq:autofunzioni} and~\eqref{eq:parametri} 
it is clear that only $d$ and $q$ determine the
form of the eigenfunctions $p_\nu$; $a$ and $b$ are scale factors for the
spectrum of $\mathscr{L}_{\text{\sc fp}}$ and for the variable $R$. 
The Gauss hypergeometric function is analytic 
into the complex plane with a cut along the positive real axis from~1 
to~$\infty$~\cite{Copson}. 
The eigenfunctions~\eqref{eq:autofunzioni} are then analytic
in their entire domain of definition~$(0,\infty)$.

\subsection{Discrete branch of the spectrum}
\begin{figure}[h]
\centering
\includegraphics[height=5cm]{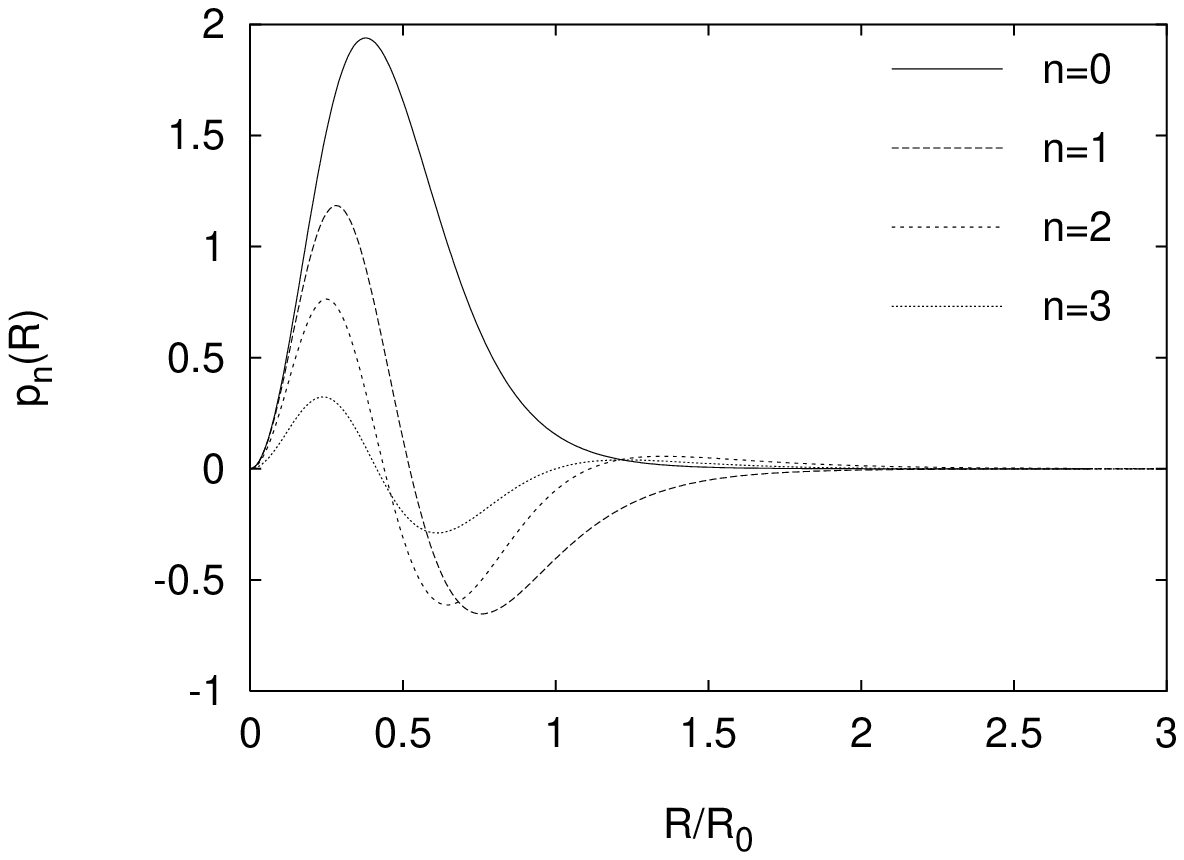}
\caption{}
\label{fig:3}
\end{figure}

The discrete spectrum of the operator $\mathscr{L}_{\text{\sc fp}}$
is defined by the first of conditions~\eqref{eq:ortog}.
From the formula expressing the analytic continuation of the hypergeometric
series in the neighborhood of infinity 
(see~\cite{Gradshteyn} formulas 9.132.2 and 9.154-5) it is 
easily shown that $p_\nu$ is normalizable only when $\gamma-\alpha_\nu$ is a
non-positive integer. Hence, the discrete levels are 
$$\nu_n=2an(q-2n) \qquad n<q/4$$
For a fixed positive $q$ the set of discrete eigenvalues is bounded from
above by $aq^2/4$ and consists of a finite number of levels.
If $0<q<4$, the only discrete eigenvalue is $\nu=0$, associated with the 
stationary pdf; with increasing $q$ more levels appear (see fig.~\ref{fig:3}). 
If $q$ is negative, the spectrum of $\mathscr{L}_{\text{\sc fp}}$ has not a discrete
branch and the pdf of the elongation does not tend to
a stationary limit.

From Euler's formula the eigenfunctions associated with the discrete part of the
spectrum 
are rational functions (see~\cite{Gradshteyn} formula 9.131 and~\cite{Goursat}
pp.~46-48):
\begin{multline}
\label{eq:autof_discr}
p_{\nu_n}(R)=N_n\, R^{d-1}\left(1+\frac{a}{b}R^2\right)^{-(d+q)/2} 
\\
\times\sum_{m=0}^n
\frac{(-n)_m \,(n-q/2)_m \,(-a/b)^m}{(d/2)_m \,m!}\,R^{2m}
\qquad (n<q/4)
\end{multline}
where $(A)_m$ denotes the Pochhammer symbol
$(A)_m=
A(A-1)\dots(A+m-1)$, $(A)_0=1$ 
and $N_n$ is the normalization constant (see~\cite{Gradshteyn} formula 3.194.3):
\begin{multline*}
N_n=
\frac{2\left(\frac{a}{b}\right)^{d/2}\Gamma\left(\frac{q+d}{2}\right)}%
{\left[\Gamma\left(\frac{q}{2}\right)
\Gamma\left(\frac{d}{2}\right)\right]^{1/2}}
\,\Big[
\sum_{m=0}^n\sum_{\ell=0}^n
\Gamma\left(\textstyle{\frac{d}{2}}+m+\ell\right)
\Gamma\left(\textstyle{\frac{q}{2}}-m-\ell\right) \\[0.2cm]\times
\frac{(-1)^{m+\ell}(-n)_m\, (-n)_\ell\, (n-q/2)_m\, (n-q/2)_\ell}{%
(d/2)_m \, (d/2)_\ell\, m!\, \ell!}\,
\Big]^{-1/2}\end{multline*}
For large $R$ the eigenfunctions $p_{{\nu_n}}$
have a power law decay
with an exponent depending on the order $n$
$$
p_{\nu_n}(R)=O( R^{-1-q+2n})\qquad \text{as} \ R\to\infty
$$

\subsection{Continuous branch of the spectrum}
The continuous spectrum consists of all real values $\nu\in(aq^2/4,\infty)$.
The eigenvalues can then be written in the form 
$\nu=aq^2/4+4au^2$, $u$ being a positive real number, and  for the sake of 
simplicity the orthogonality condition~\eqref{eq:ortog} may be replaced
by
\begin{equation}
\label{eq:ortog_new}
\int_0^\infty p_u(R)p_{u'}(R)[p_0(R)]^{-1}dR=\delta(u-u')
\end{equation}
Thus, the eigenfunctions associated with the continuous branch of the spectrum 
are 
\begin{equation}
\label{eq:autof_cont}
\textstyle
p_u(R)=N_u\, R^{d-1} F(\frac{d}{2}+\frac{q}{4}-iu,
\frac{d}{2}+\frac{q}{4}+iu,\frac{d}{2},-\frac{a}{b} R^2)
\end{equation}
where the normalization coefficient can be computed from the asymptotical
expression of the hypergeometric function (see~\cite{Gradshteyn} 
formula 9.132.2 and~\cite{Landau}):
$$
N_u=\frac{2\sqrt{N_0}}{\sqrt{\pi}\,\Gamma\left(\frac{d}{2}
\right)}\left(\frac{a}{b}\right)^{d/4}\left| 
\frac{\Gamma\left(\frac{d}{2}+\frac{q}{4}+iu\right)
\Gamma\left(-\frac{q}{4}+iu\right)}{\Gamma\left(2iu\right)}\right|.
$$
For large $R$ the eigenfunctions $p_u$ are oscillating functions with
amplitude decaying as $R^{-1-q/2}$:
$$
p_u(R)\approx \frac{2}{\sqrt{\pi}}\,
\left(\frac{a}{b}\right)^{-(d+q)/4}\sqrt{N_0}\,
\frac{\cos(2u\ln R+\delta_u)}{R^{1+q/2}} \qquad \text{as} \ R\to\infty
$$
The phase $\delta_u$ reads
$$
\delta_u=\arg \left[\frac{\Gamma\left(2iu\right) \, 
\left(\frac{a}{b}\right)^{iu}}%
{\Gamma\left(\frac{d}{2}+\frac{q}{4}+iu\right)
\Gamma\left(-\frac{q}{4}+iu\right)}\right]
$$
Assuming the completeness of the set of eigenfunctions~\eqref{eq:autof_discr} 
and~\eqref{eq:autof_cont}, we conclude that the transition probability 
of the elongation of an elastic dumbbell transported by a velocity field 
with the Batchelor-Kraichnan statistics can be written in the form
\begin{multline}
\label{eq:expansion}
p(R,t|\rho,0)=
\sum_n^{q/4-1}[p_0(\rho)]^{-1}p_{\nu_n}(\rho)p_{\nu_n}(R)e^{-\nu_nt}+\\
\int_0^\infty du [p_0(\rho)]^{-1}p_u(\rho)p_u(R)e^{-\nu(u)t}
\end{multline}

\section{Moments of the elongation}
\label{app:B}
From the Fokker-Planck equation~\eqref{eq:FP} we can derive by direct 
integration a hierarchy of ordinary differential equations for the 
moments of the elongation $y_m(t)\equiv\langle R^m\rangle(t)$:
\begin{equation}
\label{eq:momenti}
\dot{y}_m=am(m-q)y_m+bm(d+m-2)y_{m-2}
\end{equation}
with initial condition $y_{m}(0)=y^0_m$. Since we are interested
in the long time intermittency of $p(R,t)$, we restrict attention to 
even-order moments ($m=2i$ with integer $i$). 
In this case the hierarchy of equations~\eqref{eq:momenti}
is closed by the normalization condition $y_0=1$ and the $2i$-th order
moment is the solution of a system of $2i$ linear differential equations.
The solution can be obtained by Laplace transformation and direct matrix 
inversion\footnote{Odd-order moments satisfy the same hierarchy 
of equations, but the problem has an infinite number of dimensions. 
This case, therefore, is more delicate and would require studying 
the convergence of an infinite series~\cite{Graham_82}.} 
(see, e.g., ref.~\cite{Graham_82} for a description of the method). 
Define the Laplace transform of the $2i$-th order moment as
$$
\tilde{y}_{2i}(s)=\int_0^\infty e^{-st}y_{2i}(t)dt
$$
Equation~\eqref{eq:momenti} can be rewritten in the form
$$y^0_{2i}=(s-g_{2i})\tilde{y}_{2i}-h_{2i}\tilde{y}_{2i-2}$$
where 
$$g_i=ai(i-q) \qquad\text{and}\qquad h_i=bi(d+i-2)$$
or in matrix form
\begin{equation}
\label{eq:Laplace}
y^0_{2i}= \sum_{j=0}^i A_{ij}\tilde{y}_{2j}
\end{equation}
with
$$A_{ij}=(s-g_{2i})\delta_{ij}-h_{2i}\,\delta_{i,j+1}$$
By matrix inversion we obtain
$$
A_{ij}^{-1}=
\begin{cases}
\displaystyle
\dfrac{1}{h_{2j}}\prod_{k=j}^i \dfrac{h_{2k}}{(s-g_{2k})}
& i\ge j\\
0 & i<j
\end{cases}
$$
Equation~\eqref{eq:Laplace} can then be solved for
$\tilde{y}_{2i}$:
\begin{equation}
\label{eq:inv_Laplace}
\tilde{y}_{2i}(s)=\sum_{j=0}^i\dfrac{y^0_{2j}}{h_{2j}}
\prod_{k=j}^i \dfrac{h_{2k}}{(s-g_{2k})}
\end{equation}
Inverting the Laplace transform by
$$
y_{2i}(t)=\dfrac{1}{2\pi i}\int^{B+i\infty}_{B-i\infty}
e^{st}\tilde{y}_{2i}(s)\,ds
$$
we finally obtain an explicit expression for the $2i$-th order moment
as a finite sum (see ref.~\cite{Erdelyi} for the inversion of 
eq.~\eqref{eq:inv_Laplace}):
\begin{equation}
\label{eq:y2i}
y_{2i}(t)=\sum_{j=0}^i y^0_{2j}
\sum_{\ell=j}^i
e^{g_{2\ell}t}\dfrac{\prod_{k=j+1}^i h_{2k}}{f_\ell(g_{2\ell})}
\end{equation}
where
$$
f_\ell(s)=\dfrac{f(s)}{s-g_{2\ell}}
\quad\text{and}\quad
f(s)=\prod_{k=j}^i(s-g_{2k})
$$
In the stretched state,
the exponent $g_{2i}$ is an increasing function of $i$; therefore
even-order moments grow asymptotically like $e^{g_{2i}t}$.
In the coiled state,  all convergent moments converge exponentially
to the stationary value~\eqref{eq:Rm} 
like $e^{g_{2}t}$; divergent moments
behave asymptotically as in the stretched state.

\section*{Acknowledgements}

We are grateful to M.~Chertkov, J.~Davoudi,
E.~De Vito, B.~Eckhardt, J.~Schumacher, E.~Massa, A.~Maz\-zi\-no, 
L.~Ro\-sas\-co, V.~Steinberg and M.~Vergassola for helpful suggestions. 
We would like to thank R.~Graham
for pointing out to us ref.~\cite{Graham_82}.
This work was supported in part
by the European Union under contracts No. HPRN-CT-2000-00162 and 
HPRN-CT-2002-00300, and by the Italian Ministry of Education, University
and Research under contracts Cofin 2001 (prot. 2001023848) 
and  Firb SMFIRPPAR.

\newpage

\begin{center}
FIGURE CAPTIONS
\end{center}

Fig. 1: Elastic dumbbell model. The polymer is modeled by two
beads connected by a linear spring. Each bead experiences
the hydrodynamic drag force and thermal noise.

\vspace*{0.4 cm}

Fig. 2: Batchelor-Kraichnan model: stationary pdf of polymer elongation
in three-dimensions; linear scale (left) and power-law
tails in logarithmic scale (right). 
With increasing Weissenberg number, polymers get more and more elongated.

\vspace*{0.4cm}

Fig. 3: Batchelor-Kraichnan model: pdf of polymer elongation in the 
stretched state ($\text{Wi}=1.5$, $d=3$) in linear scale on the left and 
logarithmic scale on the right.
The pdf is intermittent in 
time and deviates increasingly from a Gaussian distribution.  

\vspace*{0.4cm}

Fig. 4: Two-dimensional Navier-Stokes flow:
the stationary pdf of polymer elongation for $\text{Wi}=0.4$.
The straight lines denote the power laws $R$ (left) and $R^{-1-q}$ with
$q=1.32$ (right).

\vspace*{0.4cm}

Fig. 5: Left: snapshot of the vorticity field 
$\omega={\bm \nabla}\times {\bm v}$
for two-dimensional Navier-Stokes turbulence. Right: Polymer end-to-end separation
${\bm R}$ at the same time as in the left panel.

\vspace*{0.4cm}

Fig. 6: Two-dimensional Navier-Stokes flow:
pdf of polymer elongation in the stretched state ($\text{Wi}=1.6$)
for different times $t/\tau=1,2,3,4$.

\vspace*{0.4cm}

Fig. 7: Batchelor-Kraichnan model:
eigenfunctions associated with the discrete part of the spectrum of
the Fokker-Planck operator for $q=13$, $a=b=1$, $d=3$.

\newpage

\newpage                       

\newpage

\newpage

\newpage

\newpage

\newpage

\end{document}